\documentclass[sigconf]{acmart}
\AtBeginDocument{%
  \providecommand\BibTeX{{%
    \normalfont B\kern-0.5em{\scshape i\kern-0.25em b}\kern-0.8em\TeX}}}

\usepackage{xcolor}

\newcommand{\change}[1]{{\textcolor{black}{ #1}}}

\copyrightyear{2024}
\acmYear{2024}
\setcopyright{acmlicensed}\acmConference[CHI '24]{Proceedings of the CHI
Conference on Human Factors in Computing Systems}{May 11--16,
2024}{Honolulu, HI, USA}
\acmBooktitle{Proceedings of the CHI Conference on Human Factors in
Computing Systems (CHI '24), May 11--16, 2024, Honolulu, HI, USA}
\acmDOI{10.1145/3613904.3642754}
\acmISBN{979-8-4007-0330-0/24/05}

\usepackage{soul}

\begin{document}   

\title[Gulf of Envisioning]{Bridging the Gulf of Envisioning: Cognitive Challenges in Prompt Based Interactions with LLMs}





\author{Hari Subramonyam}
 \affiliation{%
  \institution{Stanford University}
   \country{USA}
 }
 \email{harihars@stanford.edu}

 \author{Roy Pea}
 \affiliation{%
  \institution{Stanford University}
   \country{USA}
 }
 \email{roypea@stanford.edu}

 \author{Christopher Lawrence Pondoc}
 \affiliation{%
  \institution{Stanford University}
   \country{USA}
 }
 \email{clpondoc@stanford.edu}

 \author{Maneesh Agrawala}
 \affiliation{%
  \institution{Stanford University}
   \country{USA}
 }
 \email{magrawala@stanford.edu}

 \author{Colleen Seifert}
 \affiliation{%
  \institution{University of Michigan}
   \country{USA}
 }
 \email{seifert@umich.edu}

\renewcommand{\shortauthors}{Subramonyam, et al.}

\begin{abstract}
Large language models (LLMs) exhibit dynamic capabilities and appear to comprehend complex and ambiguous natural language prompts. However, calibrating LLM interactions is challenging for interface designers and end-users alike. A central issue is our limited grasp of how human cognitive processes begin with a goal and form intentions for executing actions, a blindspot even in established interaction models such as Norman's gulfs of execution and evaluation. To address this gap, we theorize how end-users `envision' translating their goals into clear intentions and craft prompts to obtain the desired LLM response. We define a process of \textit{Envisioning} by highlighting three misalignments on not knowing: (1)  what the task should be, (2) how to instruct the LLM to do the task, and (3) what to expect for the LLM’s output in meeting the goal. Finally, we make recommendations to narrow the gulf of envisioning in human-LLM interactions. 
\end{abstract}



\begin{teaserfigure}
  \includegraphics[width=\textwidth]{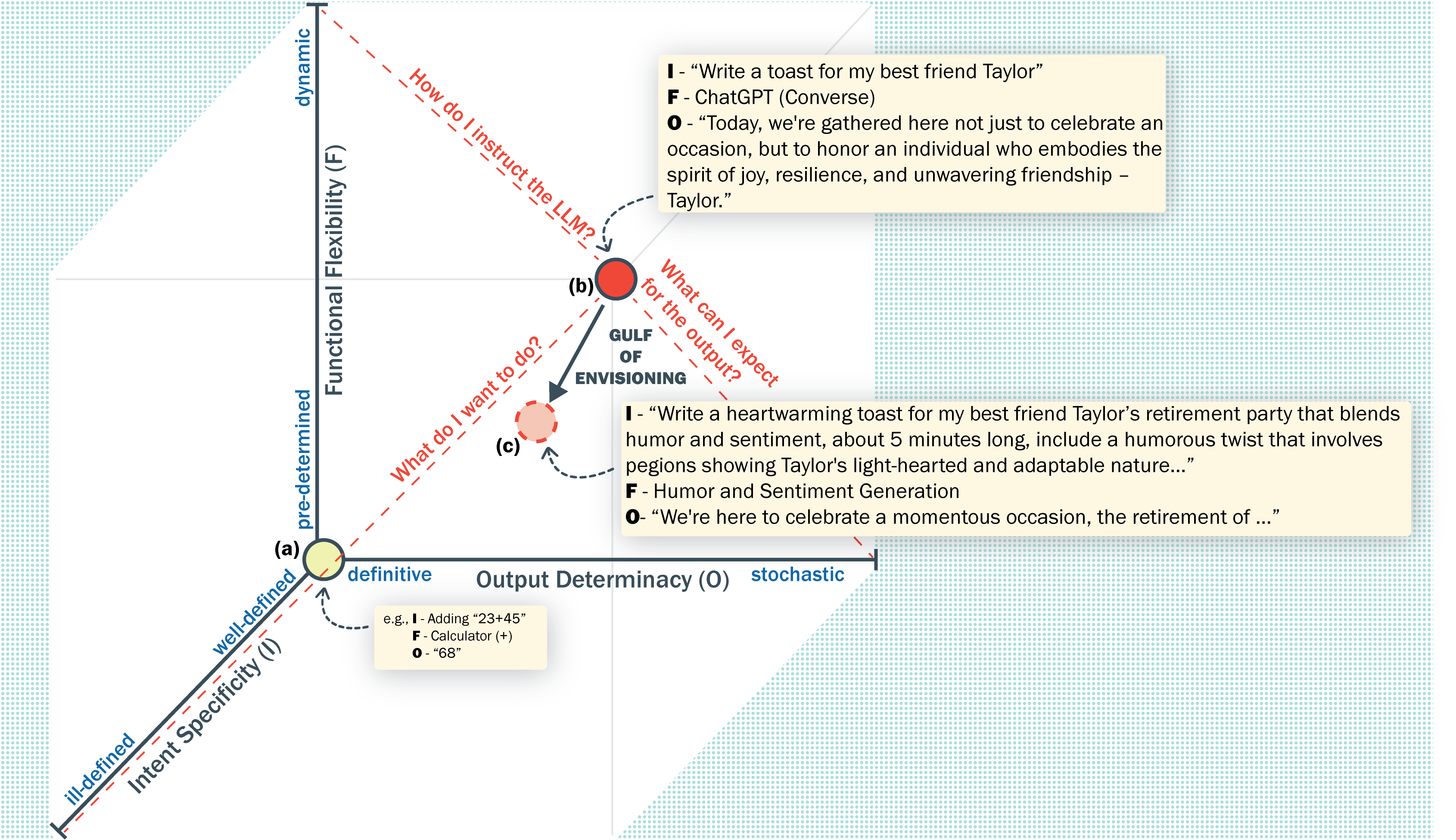}
  \caption{A conceptual framework for LLM-based Interactions along three dimensions: (1) Intent Specificity, (2) Functional Flexibility, and (3) Output Determinacy. Point (a) indicates conventional interfaces such as a Calculator with pre-determined functionality and affordances for interaction. Point (b) represents conversational LLMs such as ChatGPT with dynamic functionality. Gulf of Envisioning is the challenge users face in formulating prompts to generate high-quality outputs, i.e., point (c). }
  \Description{Dimensions of interactions}
  \label{fig:teaser}
\end{teaserfigure}






\maketitle

\section{Introduction}

\change{Large language models (LLMs) such as ChatGPT~\cite{chatgpt} have demonstrated remarkable capabilities in generating content that is novel, coherent, and contextually relevant. These models can perform a wide array of tasks, from writing comprehensive essays to creating artwork and even producing functional software interfaces, showcasing a high degree of creativity and adaptability. However, they also require careful \textit{guidance} to ensure the generated content is appropriate and in alignment with human goals and intentions. For instance, if an end-user wishes to leverage an LLM to craft a toast and prompts the LLM with \textit{``Write a toast for Taylor''}, the output may be incomplete without providing the desired qualities. The human must be more specific about their \textit{intentions} (such as, ``Write a heartwarming toast for my best friend Taylor's retirement party, about 5 minutes long, include a humorous twist, and wish them well on the golf course''). Formal and anecdotal evidence (e.g.,~\cite{zamfirescu2023johnny, kocon2023chatgpt, kaddour2023challenges, agrawalablog}) suggests that effectively prompting LLMs to produce outputs similar to human-generated content remains challenging. If intentions are expressed too vaguely or lacking specific detail, the LLM may generate responses that are generic, irrelevant, or off-topic~\cite{holtzman2021surface,zamfirescu2023johnny, kim2023understanding}. Iterating with an LLM can correct and progressively guide generation, but playing a ``20-questions'' or ``Hot or Cold'' guessing game may be inefficient for longer output and lead to a local minima within the solution space~\cite{simon1971human}. Further, humans show \textit{fixation} on initial examples that interfere with exploring alternative solutions~\cite{jansson1991design, leahy2020design}. \textbf{In this work, we draw from theories across HCI and cognitive science to characterize the nature of the cognitive challenges for humans in dialogic interactions with intelligent generative agents.}}

\change{As shown in Figure~\ref{fig:teaser}, the shift towards LLM-powered interfaces can be characterized along the following three dimensions: (1) Functional Flexibility, (2) Intent Specificity, and (3) Output Determinacy. The vision for artificial general intelligence (AGI) includes LLMs with a robust theory of mind (ToM) of humans (and vice-versa), and would allow effective collaboration across numerous tasks. While not yet at this level of general intelligence~\cite{morris2023levels, ullman2023large}, current LLMs do exhibit \textit{dynamic} capabilities in that they are able to fulfill a broad range of tasks and generate ad-hoc functionality in response to prompt inputs (e.g., ``rewrite these appliance installation instructions for a five-year-old''). This flexibility contrasts with conventional direct manipulation interfaces with pre-determined functionalities (e.g., a calculator). Even contemporary natural language interfaces, while aiming for linguistic variability and conversation-style interactions, remain essentially function-specific, like locating the closest coffee shop~\cite{folstad2017chatbots}, updating specific attributes of vector graphics~\cite{laput2013pixeltone}, controlling task workflows~\cite{subramonyam2018taketoons}, or constructing data charts~\cite{setlur2016eviza}.} 

\change{Second, LLMs exhibit open-ended generative characteristics through the vast quantity of linguistic patterns and information learned during training. Their capacity to generate diverse responses to a given input underscores their utility in creative and conversational tasks. While LLMs can produce determinate, correct solutions to closed-ended problems, they offer great value by generating many different solutions to an open-ended problem~\cite{newell1972human}. As indicated along the \textit{Output Determinacy (O)} dimension in Figure~\ref{fig:teaser}, the LLM's capacity to generate varied and unexpected output defines its inherent unpredictability. Humans need to provide oversight and guidance by validating facts, verifying relevance, checking for biases, and evaluating output quality. Third, end-users can converse with LLMs through prompts expressing goals and intentions at any level of specificity. As indicated in the \textit{Intent Specificity (I)} dimension in Figure~\ref{fig:teaser}, structure and convention in the input are not imposed by the LLM, leaving end-users uncertain about how to formulate input to improve LLM generation. The new capabilities of LLMs greatly extend human-machine interactions along these three dimensions into new territory through an interaction model where (1) operational scope is not restricted to pre-programmed tasks, (2) all input intentions are allowed with an ``anything goes'' approach, and (3) outputs are probabilistic rather than determinate.} 

In this work, we examine the transformative impact of generative AI systems for human-machine interaction, focusing on how this shift from conventional interfaces alters the design and usability of interactions on these three dimensions. Hutchins et al.~\cite{hutchins1985direct} offer a model of interface-design-challenges in software systems, including an ``execution gulf'' between user intentions and system actions and an ``evaluation gulf'' between system output and user understanding of its genesis. Their general principle is that as the \textit{distance} between the human's intentions and the system's interface increases, the costs of interaction increase. LLMs transform human-machine interaction to substantially reduce this distance, and therefore the costs, by defining interaction as human formulation of intentions through natural language dialogue leading to desired output~\cite{zamfirescu2023johnny, ko2004six}. LLMs narrow the gulf of execution by eliminating conventional needs for \textit{action specification} and \textit{execution} \cite{hutchins1985direct}, leaving only \textit{intention} to the user. However, the gulf of evaluation may increase by challenges to \textit{perceive}, \textit{interpret}, and \textit{evaluate} output \cite{hutchins1985direct} given the LLMs' probabilistic process. 

What are the consequences of these new LLM features enabling success on complex tasks -- flexibility in functional scope, variation in intention specificity, and probabilistic processes and outputs -- on the nature and costs of human interaction? We suggest this new LLM interaction process poses new challenges for people, which we call, ``\textit{the gulf of envisioning}.''  Concretely, the gulf of envisioning characterizes the \textit{distance} between the human's initial intentions and their formulation of a prompt that foresees how LLM capabilities and training data can be leveraged to generate high-quality output. Envisioning includes at least three challenges for humans interacting with LLM systems: (1) how to set my goals and intentions such that the LLM can accomplish the task -- \textit{the capability gap}, (2) how to best instruct an LLM about my goals (i.e., prompt engineering) -- \textit{the instruction gap}, and (3) what to expect for the LLM's output -- \textit{the intentionality gap}.   
  
In this paper, we formulate a new model of interaction for human-LLM interfaces in which \textit{intentions are the actions.} Our key contributions include (1) a characterization of how transformative LLM natural language interfaces yield both expansive functionality and new challenges in bridging intentions and outcomes; (2) an updated model of human-machine interaction identifying the process of envisioning execution; and (3) a set of design patterns and guidelines for human-LLM interfaces along with an analysis of interfaces for three types of generative tasks.

\section{Intentions and Interactions in Conventional Software Systems} \label{sec:paradox}
A primary focus of HCI is designing interfaces that mediate the \textit{interactive} relationship between an end-user and a computational system to accomplish a human goal. To this end, researchers and practitioners have conceived several different interaction paradigms (summarized in~\cite{hornbaek2017interaction}), proposed frameworks to understand challenges in human-machine interactions~\cite{norman1986cognitive, hutchins1985direct}, and identified ways to solve those challenges through the use of affordances, feedback mechanisms, task-oriented design, etc.~\cite{wood1997user,gibson1977theory, de2005semiotic}.

\begin{figure*}[t!]
  \centering
  \includegraphics[width= 0.7\textwidth]{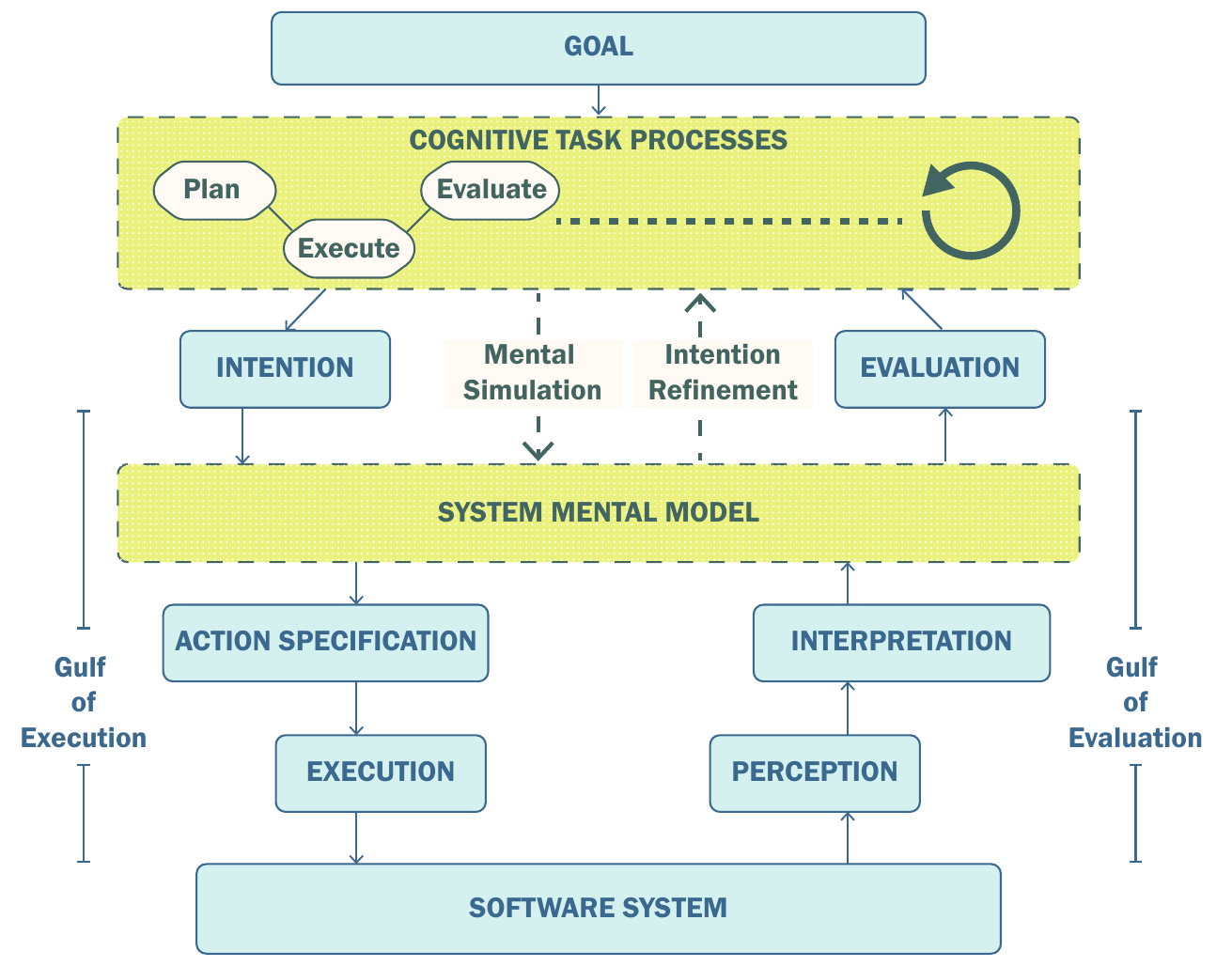}
  \caption{ Expanded View of Norman's seven-stage model of interaction. Norman's seven-stage model (blue blocks) is a valuable tool when designing interfaces. However, while the second stage -- forming the intention -- has often been overlooked in most HCI research, it is a crucial part of human-LLM interactions. We add the cognitive task and system mental blocks (in green) to illustrate the mechanism for intention formation and its interaction with action specification. }
  \label{fig:norman-model}
\end{figure*}

However, an intriguing question remaining underexplored is \textbf{how end-users conceive intentions when engaging with the interface}. In Norman's seven-stage model~\cite{norman1986cognitive}, interaction consists of (1) Establishing the Goal, (2) Forming the Intention, (3) Specifying the Action Sequence, (4) Executing the Action on the System's Interface, (5) Perceiving the System's State as a Response to the Action, (6) Interpreting the State, and (7) Evaluating the System State with respect to the Goals and Iterating until the goal is achieved (see Figure~\ref{fig:norman-model}). In much of HCI work, stage two -- \textit{forming the intention} -- is assumed as given~\cite{hornbaek2017interaction}. For instance, when cutting and pasting a paragraph of text in a word processor or clicking on the `Bold' font button, how much do we know (or care) about the underlying intentions leading a user to execute those actions? We posit that while this gap from goal to intention has been inadvertently bypassed in traditional design approaches, it emerges as a critical challenge that must be addressed in human-LLM interactions. In this section, we explore this overlooked aspect of intention formation during interactions and postulate its role in LLM-powered interfaces. 

\subsection{Defining Intentions}
At the highest level, an intention is a \textit{situated pursuit} of a goal that is attainable through the \textit{execution of a process}~\cite{jackson1995software, bertelsen2003activity} of a certain sequence of actions conceived as leading towards a goal. An intention is an intermediate cognitive state that translates the abstract desire (goal) into concrete actions. Research in cognitive task analysis suggests that intentions are not just spontaneously generated but arise from a foundation of knowledge, thought processes, and goal configurations~\cite{clark2008cognitive}. Intentions include aspects such as declarative knowledge (understanding ``what'' needs to be done), procedural knowledge (knowing the ``how-to'' of a task), decision points (key moments in reaching a goal where decisions are necessary), and cognitive skills (the mental abilities required to carry out the task). Further, intentions are tied to goals through complex hierarchical structures, and they emerge as the user works towards achieving those goals. This key cognitive science insight was advanced in seminal works by Karl Lashley ~\cite{lashley1951problem} and Miller et al.~\cite{miller1960plans} in their focus on the hierarchical structure of nested subroutines in human action, opposing behaviorist conceptions of sequential behavior as a chain of stimulus-response associations, and it is further elaborated in current work in cognitive and computational neuroscience ~\cite{botvinick2008hierarchical}.

\subsection{Role of Intentions in Interactions} 
How does a user formulate intentions and then specify them as actions to be executed by a system? Let's start with the more familiar aspect of this question, which is action specification, before delving into intentions. According to Norman, users have (or rather acquire) in their mind, mentally represented models of the target system that provide them with the ``predictive and explanatory powers'' to understand interactions~\cite{norman2014some}. We refer to these representations about systems that \textbf{help with action specification as system mental models}. These models primarily comprise knowledge about how the system operates, including its constituent parts and their interrelations, their inherent processes, and their impact on the system output~\cite{carroll1988mental}. While many researchers have characterized the nature of system mental models (e.g., ~\cite{young2014surrogates, bayman1984instructional, mayer1981psychology, moray1987intelligent}), the common purpose pertaining to HCI is to equip users to determine which action to execute by allowing them to mentally \textit{simulate} the action~\cite{staggers1993mental}. For instance, imagine that a writer (the user) is about to type a long and detailed section header. Before committing, they run a \textit{mental simulation} using the system mental model: they envision typing the header in full length, foreseeing it might take up too much space or look cumbersome in the document. Consequently, they may consider changing the header size or shortening the text. This \textit{mental rehearsal} helps them anticipate potential readability issues or aesthetic concerns and select their action sequence accordingly. 

In the above example, intention formation has not been considered. That is, when mentally simulating the typing of the header, how did the user conceive the header text (i.e., input to the system mental model) in the first place?  Of course, this pertains to the user's goal; let us assume it is to produce a Wikipedia article on Chocolate. A specific intention is the user's decision or plan to type a long, detailed section header in the document, say \textit{``Ethical and environmental implications of cacao bean farming in various tropical regions and their socio-economic impact.''} The formulation of this intention is tied to the underlying \textit{cognitive task processes} of writing that are independent of the technological system. Generally, they concern how people accomplish mental tasks through component processes, such as retrieval from memory, transforming and combining ideas, and using procedures, etc. Specific to writing, such a cognitive task process is formulated by Hayes and Flower, and it involves three \textit{intertwined} cognitive stages, including planning, translating, and reviewing~\cite{hayes2013new}. The specifics of this model are less important at this point, and we will discuss cognitive task processes in detail in Section~\ref{sec:model}. But what it is important to understand is that these \textbf{cognitive task processes support intention formation.} Even for a simpler goal, such as setting a wake-up alarm, the end user must mentally determine the time they need to leave for work, the time it might take to get ready, plan their commute, etc. This is all part of the cognitive task processes. 

Critically, these two stages -- formulating intentions through cognitive task processes and specifying actions through system mental model simulations -- \textit{interact}. In order to generate their intentions for a given goal, the user must be engaged in the cognitive processes needed for that goal. To specify the actions for the system to take based on formulated intentions, the user must simulate the system's mental model to determine how selected actions may influence the execution of their intention. In reality, users constantly adapt their intentions based on both their dynamic cognitive task processes (what they want to accomplish) and their system mental model (what they believe can be accomplished through system actions). In the Wikipedia example, the mental simulation of typing a long header may result in the \textit{reformulation} of the intention to a shorter title, say ``Ethical impacts of Cacao farming.'' Thus, interaction requires (1) intention (formulation and refinement), (2) system mental models that allow a cognitive walkthrough of how the intention might play out at the end of the interaction, and (3) the ability to mentally project the results of their intended actions (i.e., anticipate the eventual outcome of the interaction).

\subsection{Role of Design in Mapping Intentions to Actions}
Until LLMs, HCI approaches have successfully \textbf{interwoven intentions and actions} during interaction design. Through human-centered design approaches, interaction designers strive to ensure that every potential action available on an interface aligns intuitively with the users' goal-based  intentions~\cite{wood1997user, norman2014some}. Even for complex cognitive goals such as reading, writing, reasoning, and sense-making, researchers have developed interactive interfaces by understanding the underlying cognitive process model (e.g.,~\cite{kuznetsov2022fuse, subramonyam2020texsketch,endert2015semantic}). For instance, in texSketch~\cite{subramonyam2020texsketch}, they build interactions around the cognitive processes in reading, such as selection, organization, and integrated comprehension. This alignment fosters a symbiotic relationship where users feel the system is an extension of their thought processes, making the interaction feel natural and efficient. 

Second, design also plays a critical role in helping users to formulate a mental model of the system, and through it, learn to anticipate system outcomes. In conventional HCI, designers invent a \textit{conceptual model} -- an abstract representation or framework to communicate how a system operates -- based on human-centered design practices~\cite{norman2014some}. For instance, the designer may analyze users' past experiences and prior knowledge and draw from familiar analogies and metaphors (e.g., the desktop metaphor)  to shape the presentation and behavior of the system~\cite{neale1997role}. According to Hutchins~\cite{hutchins1987metaphors}, conceptual models function as metaphors on several levels. At the broadest level, they're shaped by the users' primary goals. In a word-processing tool, this could mean a ``blank page'' metaphor aligning with users' primary goal of creating a document. Moving in more granularly, metaphors at the interaction level inform users about their computer interactions and often remain consistent across tasks (e.g., a ``clipboard'' for copying and pasting). Lastly, metaphors at the task domain level provide an understanding of how tasks are organized and structured, such as the use of tabs or headers and footers to indicate document structuring during writing. By establishing this foundation by means of HCI design, users can more readily form a coherent internal representation of the system's operations and functionalities. This representation enhances their ability to plan and execute complex tasks while seamlessly aligning their goal intentions with the appropriate actions. 

With LLM systems, the \textbf{link between user intentions and system actions is less clear, and end-users lack adequate mental models of LLMs.} Consequently, LLM interactions become challenging for users. In the next section, we postulate how these constraints challenge forming intentions for interactions.

\section{Envisioning Intentions in LLM Interactions}\label{sec:model} 
LLMs represent a significant leap forward in the evolution of natural language processing (NLP) capabilities. From the perspective of interactive system design, LLMs obviate the need for structured interfaces with preset actions for implementing intentions in favor of \textit{unconstrained} use of natural language (note that we address specific interface designs for LLMs in Section~\ref{sec:cases}).  With a dynamic operational scope lacking explicit interface actions, how do users formulate their intentions and then express them as prompt inputs? One proposed solution is to treat LLMs as if they are human and engage in a conversation with them~\cite{reeves1996media, dillion2023can, cui2023drive, zamfirescu2023johnny}. Here, we examine this approach to intention formation by considering a specific generative task for LLMs -- writing -- in three aspects of human-to-human interactions: models of communication, roles and expertise, and theory of minds.  

\subsection{Cognitive Processes in Generative Tasks}
\textbf{What happens when humans perform generative tasks?} Studies show that human performance of `generative' tasks, such as writing, creating new ideas, coding, and reasoning, appears endlessly variable. At a high level, human task performance takes the form of repetitive cycles of cognitive processes where a change in process, failure, or success is not predictable. A general observation is task performance proceeds through cycles of \textit{``plan a little''} and \textit{``do a little''}~\cite{miller1956magical,decety2006power}. There is no definitive cognitive process for how to accomplish a generative task, though much of education is aimed at instruction on exactly these tasks. Models offer few distinctions among tasks despite distinctive aims, and a high degree of variation in cognitive processes is observed for the same task in different people and for the same person repeating a task.   

\change{The predominant model of generative tasks is Newell \& Simon's (1972) model of problem-solving ~\cite{newell1972human}. They defined a problem space, including the current state, a desired goal state, and all available actions or operators. The process of problem-solving is defined by actions taken to bridge the gap between nearby states, describing a path toward a solution. For instance, a  means-ends analysis process identifies differences between current and desired states and selects an action to decrease their distance ~\cite{newell1961computer}. The path from each state is viewed as traversing a solution space containing all possible outcomes. Problem space models describe well-defined problems where these elements are known, and branching can identify all possible combinations of actions. This approach is successful as a model for machine problem solving~\cite{newell2007computer}, but when applied to human problem solving, deterministic solution paths with well-defined goals, operators, and evaluation of options are rare~\cite{seifert1994demystification}. Typically, great variability is observed in cognitive processes, including strategies like back-tracking from a goal to a current state, mental simulation of potential actions, and generation of novel actions.}

As a consequence, \textbf{cognitive models of generative tasks describe a more variable process} with ill-defined initial states, goal states, and available actions. For example, a four-stage model of human problem solving describes cognitive processes of 1) problem identification, 2) planning, 3) implementation, and evaluation~\cite{duncker1945problem}. \textit{Defining a problem} is far from a determinate process yet central to success.  \textit{Planning} is defined as identifying and organizing sub-goals, intentions, and actions~\cite{miller2017plans,pea1982planning} to achieve a goal.  Next, an \textit{execution} phase turns intentions into actions. In addition to ``doing'' the task, people coordinate other cognitive processes, such as monitoring for errors and making real-time adjustments based on feedback. During the \textit{evaluation} phase, users are engaged in a comparative judgment of the actions' outcome in relation to a goal state. Evaluation also includes self-regulation, metacognitive awareness, and value functions, as well as explaining errors to plan subsequent intentions and actions~\cite{brown2017metacognitive, schraw1994assessing, wittrock1989generative}. These stages are so loosely defined that the 4-stage model is ubiquitous in accounts of thinking, creativity, design, and other generative tasks~\cite{duncker1945problem, howard2008describing}. When applied to an instance of task performance, the cognitive processes are indeterminate, with the progression and order of stages varying, and stages are often independently repeated. \textit{Iteration} is assumed for all stages in these models because the cognitive processes are also defined over subtasks as needed~\cite{bamberger1983learning}. 

\subsubsection{A Cognitive Task Process for Writing Tasks}
A cognitive process model for writing offers a more specific account of the steps required for completing a task~\cite{hayes2013new}. First, the writer defines the task environment, including the audience, the purpose, and tools. Next, the act of writing involves three \textit{intertwined} cognitive stages for planning, translating, and reviewing. The \textit{planning} phase creates a mental roadmap for writing. It involves retrieving ideas or content related to the topic, articulating objectives such as the intent to inform, persuade, or entertain, structuring ideas in a coherent order, and determining how to transition from one idea to the next. The \textit{translating} stage involves navigating the roadmap and putting ideas into words by converting cognitive representations into linguistic expressions. After a segment (or an entire piece) is written, a third stage involves \textit{Reviewing} it to determine whether it aligns with initial goals, such as coherency, compellingness, and clarity. Based on this evaluation, they may choose to revise the text through restructuring, adding new information, or deleting content. Crucially, this cognitive process during a writing task is not a linear progression from one stage to the next, but an iterative, dynamic process of constant repetition, refinement, and reflection as writers respond to their ongoing assessment of their created output. In this more specific model of cognitive processing in writing texts, the same general steps (planning, implementing, and evaluating) and prominent iteration are evident. In addition, a role for continual monitoring of what is produced and how it meets the goals suggests constant oversight of the generation process.

\subsection{Generative Task Interactions with an LLM Agent}

\begin{figure*}[t!]
  \centering
  \includegraphics[width=\textwidth]{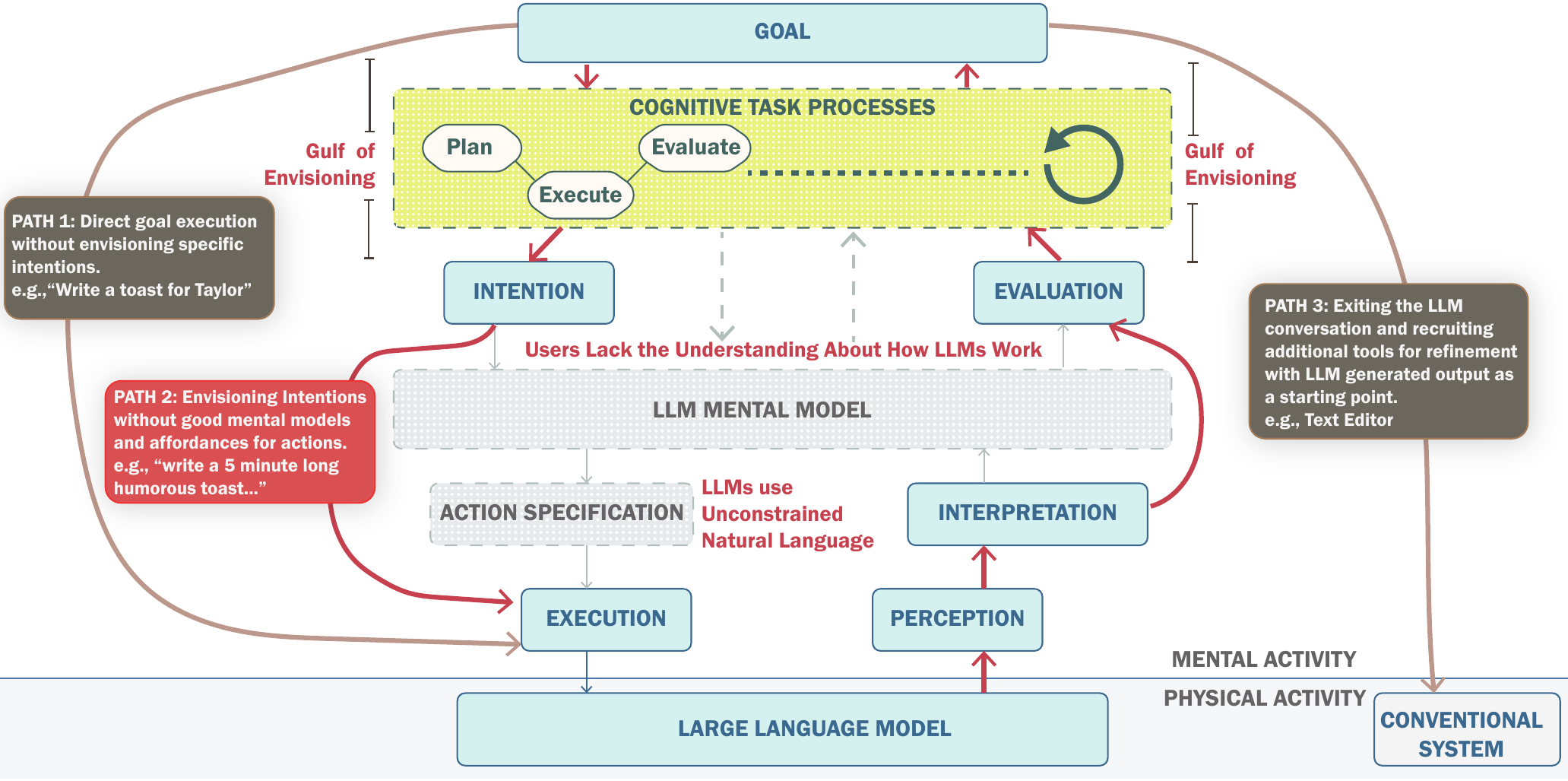}
  \caption{In the context of Norman's seven-stage model action, we highlight 
  what is missing during human-LLM interactions. Further, there are three pathways to interactions: (1) directly state their goal to the LLM, (2) formulate their intentions and provide them to the model through prompt engineering, and (3) take the LLM output and transition to a dedicated interface and system (e.g., switching from ChatGPT to a Word Processor based on an LLM generated draft).}
  \label{fig:llmmodel}
\end{figure*}

Based on this understanding of cognitive task processes, let us revisit the goal of writing a toast for Taylor's retirement. If recruiting a human to help, one might choose an expert who knows Taylor well, and has been to many such events, or an assistant with strong writing skills based on prior knowledge of the expertise required for the task and a theory of specific minds reflecting their knowledge states. Based on this understanding, we may directly ask the expert to \textit{``Write a retirement toast.''} With an LLM agent, we can assume training on everything posted on the internet, a generation process by next-word prediction, and conversational interaction to iterate on output. Based on this assumption, the most straightforward path is to simply state the goal to the LLM in their own words, as shown in Figure~\ref{fig:llmmodel}- PATH 1. Such an approach aligns with the theory of computational rationality, in which interactive behaviors fundamentally depend on the principle of expected utility~\cite{oulasvirta2022computational}. That is, users tend towards behaviors that maximize expected utility within specific constraints, called bounded rationality~\cite{simon1997models}. Constraints can emerge internally from cognitive or bodily capacities, or externally from the environment. A key element in this determination is the perception of effort required to improve the input quality and the potential gain in output quality. Given that LLMs are likely trained on databases containing toasts, the resulting output may satisfice ~\cite{simon1956rational, tidwell2010designing}, and generic or adequate answers may be readily available. 

Experienced users may learn that instead of directly stating the goal, a greater \textit{effort} to formulate intentions within prompts can increase the utility by arriving at better or quicker solutions. For example, while it is possible to iterate based on output evaluation, the time to read and reformulate prompts has costs. If the iterative steps can be combined in fewer steps, taking more care in formulating intentions may be cost-effective. For instance, in creating a prompt for the LLM to write a toast for Taylor, the user may engage in planning how they would do the task by identifying topics to include, a structure and organization, constraints for format and length, tone and writing style, desirable qualities, etc. The more developed and specific the intentions, the better the LLM's output should satisfy the user. We call this process envisioning (Figure~\ref{fig:llmmodel}- PATH 2). However, envisioning more developed intention specifications is effortful and cognitively demanding. To craft a more specific prompt for the LLM, the user must further develop their intention by mentally exploring potential plans and values as if -- but not actually -- executing them, and then adding what is discovered to the prompt. This often leads to taking the Path 1 approach without envisioning. If ``an answer'' or even a ``satisfactory answer'' is needed, envisioning may not be. But, if aiming for a high-quality answer, a user may better exploit the system's vast knowledge by envisioning possible and very desirable output.
In what follows, we discuss specific cognitive gaps in envisioning LLM interactions: 

\subsubsection{Capability Gap:}
Recall from our earlier definition that an intention includes specifications about how to execute task processes. The \textit{capability gap} concerns the users' \textbf{inability to formulate ``how to'' procedures to implement their intentions.} Defining when, where, and how one wants to take action on intentions requires added cognitive effort~\cite{PATALANO19971, seifert2001opportunism}, but has been shown to enhance the rate of successful execution~\cite{gollwitzer2020implementation}. While LLMs, with their extensive training data, can theoretically understand and generate a wide variety of bespoke task content, their very strength can also be a source of complexity for users. Take, for instance, entering a prompt for summarizing a 2-page text in 1 page. For a human writer, this task involves a myriad of choices about what to prioritize, which nuances to retain, and which details to omit. How can a user identify the right specification of intention for the LLM? 

For most generation tasks, a \textbf{cognitive process is not already well-formed} in memory but can be made more explicit (or newly generated) through planning. As described in the section on cognitive models, generative tasks are not determinate and vary with each instance of generation and each intention (every story is written with a different specific process). On the other hand,  when tasks are well practiced through experience with execution, people may have ``scripts'' or well-developed routines to complete a generative task ~\cite{schank2013scripts}. However, some tasks may be \textbf{too well learned to summarize as instructions;} for example, asking someone how to make a peanut butter and jelly sandwich without actually making it\footnote{The EXACT INSTRUCTIONS CHALLENGE video: \url{https://youtu.be/Ct-lOOUqmyY}} demands generating actions without feedback from execution. ``Doing'' offers reminders of where you take added steps to avoid errors. Without this step (since LLMs are doing the generation), users are forced to take a trial-and-error approach with differing abstractions and variations of their intentions, and then iterate based on evaluating the LLM's output. But unlike deterministic systems where a given input always produces the same output, an LLM can produce countless variations from the same input, and there is, by definition, no one correct answer in generative tasks. This \textbf{indeterminacy in LLM responses}, coupled with a\textbf{ lack of transparency in how inputs lead to LLM outputs}, can hinder users' attempts to formulate prompts.

Of course, human-LLM interactions are designed to be conversational and iterative, allowing users to evaluate output, revise prompts, and further refine intentions until satisfied. However, there are \textbf{costs associated with iteration from user experience, social exchange expectations}~\cite{cropanzano2005social}, and generative process perspectives. In envisioning intentions, users must find a balance between ``batching'' intentions in prompts and development through iteration. Intention setting in advance allows users to carefully consider their goals without potential change or interference from the LLM. For example, if writing about personal experience or emotions, iterations may draw the eventual outcome further away from the user's internal perspective. Repeated iteration over many small changes to prompts may \textbf{violate a user's conversational expectations} for ``chunking'' of information in turn-taking. Most importantly, iteration poses the \textbf{danger of fixation} on the presented output. Humans tend to become fixated on their own or others' initial proposed solutions ~\cite{ maier1931reasoning, jansson1991design, leahy2020design} and struggle with overcoming attachment to early ideas ~\cite{cross2001design}, limiting options for solutions through fixation on a particular type of idea or concept~\cite{purcell1996design}, resulting in a lack of exploration of alternatives. If a first output is viewed, the user may ``anchor'' on it as a solution and focus on local refinements to local minima, thereby missing opportunities in more distant solution spaces.  Those with less developed intentions may be more likely to show a stronger fixation on an initial LLM output due to a lack of knowledge about desirable outcomes. Experienced users may learn \textit{when} greater effort to formulate intentions within initial prompts is worthwhile in arriving at better solutions or quicker system interactions. For example, while it is possible to iterate based on output evaluation, the time to read and reformulate prompts has costs. If the iterative steps can be combined in fewer steps, taking more care in formulating intentions may be cost-effective.

\subsubsection{\change{Instruction Gap:}}
The \textit{instruction gap} refers to the user's challenges in clearly and effectively expressing their intentions in the interface as text prompts, leading to a potential mismatch between the user's intentions and what the LLM perceives and then produces. One benefit to the natural language interface for LLMs for end-users is a \textbf{wide-open input space} for specifying intentions. In writing the toast, the user's intentions as described in the text prompt can \textit{vary} across language expressions; for example, the role assumed (e.g., ``act as an event emcee,'' ``act as a best friend''), desirable qualities of output (e.g., ``make it funny,'' ``make it heartfelt''), and related text content as a starting point for generation (e.g., ``golf,'' ``Vegas''). Unfortunately, the LLMs' algorithms for learning from text corpora create a dependency on \textbf{language precision}. While human language use tolerates a wide range of expressions communicating a similar meaning, even slight changes in words can lead LLMs to produce significantly different outputs. It is possible that substantial experience with LLM use will improve a user's understanding of the precise mapping between prompt contents, at least in a domain area. For instance, more expert users may have an understanding of transformer architectures, and experienced users may know how specific keywords in the prompt (e.g., ``punchy'') influence the attention mechanism. ``Instructing'' is a much more deliberate task with individual variation, and there are few examples of product instructions done well despite their ubiquity (but see ~\cite{agrawala2003designing}). Users \textbf{challenged in performing improvisational linguistic dexterity} may not obtain satisfactory outcomes solely through trial and error iteration of prompts. Further, linguistic sensitivity makes it \textbf{difficult for users to script their intentions for the LLM upfront by anticipating potential system interpretations}. Therefore, users are likely to need a multi-turn interaction with an LLM session to determine alternative specifications for their intentions leading to better outcomes.

\subsubsection{Intentionality Gap:}
In PATH-1 interaction with the LLM, the user states their goal directly and forgoes any cognitive task processes anticipated in doing the tasks themselves  (i.e., planning what to write, executing it, and evaluating it). Instead, they start with evaluation of the LLM-generated text. Does it meet their goal? This evaluation is cognitively demanding because no ``bridging'' steps between goal and output are provided. By skipping directly from goal to evaluating output, users may lack a cohesive understanding of the task context and required content,\textbf{ hindering their ability to make comprehensive judgments about the adequacy of the output.} Drawing from philosophical terminology, users encounter an \textit{intentionality gap} while assessing texts generated by LLMs. In philosophy, intentionality refers to the capacity of mental states to be about, directed at, or represent something in the world outside of themselves~\cite{searle1983intentionality}. In the LLM context, users may not have a nuanced ``aboutness'' of the specific intentions or contexts they are considering and, therefore, may not have a clear sense of what the content should be ``about'' from their own perspective. Consequently, \textbf{without internal cognitive benchmarks of the generation process for comparison, they may become over-reliant on the LLM and avoid evaluation against thier self-generated goals.} Further, engaging in subsequent planning to identify and rectify any deviations from a desired outcome can be challenging. In other words, users producing only a goal statement will have difficulty answering questions such as, \textit{ ``Can the LLM do better?''} and \textit{``How can the LLM do better?''}

Note that this intentionality gap is different from the gulf of evaluation in Norman's model~\cite{norman1986cognitive}. With traditional HCI tasks, deterministic outcomes are easier to perceive and interpret, whether completed successfully or not. Feedback is often immediate, as with a direct manipulation user interface in which ``\ldots users can immediately see if their actions are furthering their goals~\cite{shneiderman1982future}.'' The evaluation gulf stems from ambiguous system feedback at the interface level, which may make it challenging to perceive and interpret the produced outcome. Addressing the gulf of evaluation involves improving system feedback, representation, and visibility. In contrast, the intentionality gap lies at the cognitive level due to the user's failure to create clear intentions. Bridging this intentionality gap with LLMs may require interfaces to scaffold users in creating clearer task process prompts and developing their contextualized understanding of the qualities desired in the generated text. 

\change{The idea that under-specified intentions lead to challenges in evaluation surfaces in Karl Duncker's~\cite{duncker1945problem} work on the process of finding a solution through a continual restructuring of a problem over time to develop the ``essential'' properties of a desired solution. Problems intentionally left open-ended, as in many creative tasks, are termed ``ill-defined''~\cite{simon1971human} in that \textit{incomplete} information is provided about the problem. For example, design intentions require a great deal of construction by the designer ~\cite{restrepo2004problem} to identify valuable qualities of potential solutions. Identifying the nature of a problem is key to solving problems,  creative work, and design thinking ~\cite{hargadon2006collections, dorst2011core, ecker1963artistic}. Exploring problems provides perspectives on values important in solutions~\cite{ dorst2001creativity, schon1983becoming}. The process of developing intentions helps one learn to ``see'' the desirable attributes of non-existent outcomes~\cite{dorst2011core} and specify important solution qualities. The power of exploring intention is illustrated by an early study of fine artists asked to create still-life drawings: those showing ``discovery-oriented'' behavior (rearranging objects, changing perspectives, touching objects) produced works with greater originality and higher quality (as judged by experts), and even experienced greater professional recognition and income years later. \textbf{Those who showed more consideration of the qualities of their intended piece produced better outcomes than those focused on execution}~\cite{csikszentmihalyi1971discovery, csikszentmihalyi1988creativity}.}

\subsection{Recruiting New Pathways for Interacting With LLMs}
In the Intentionality Gap, users interact directly with the LLM in evaluation mode. This means they're often trying to make judgments about the generated output and whether changes to the input may produce different outputs more aligned with their goal. \textbf{With LLMs, they must attempt these cognitive maneuvers without a foundational context or a cohesive mental model of the LLM functions.} In a third proposed interaction pathway (Figure~\ref{fig:llmmodel}- PATH 3), we suggest a dedicated interface tool with support for the user to further develop intentions and actions to more clearly direct the LLM toward their task goals.  This shift to an interface tool that knows specific complex generative tasks -- how to write a screenplay or code in Java -- facilitates the user's identification of a functional basis for the task process, enabling the user to more precisely specify, calibrate, and synchronize their intentions with corresponding plan descriptions. Critically, \textbf{the scaffolding provided by such a ``prompt prompter'' can facilitate prompt entry and completeness }by identifying expected elements of task functions that are missing in a user prompt and requesting needed information (e.g., ``who is your audience?''). Another alternative pathway is integrating LLMs within existing functional tools, as we will see in section~\ref{sec:cases}. The use of a specialized interface tool can help scaffold human interactions with LLMs by dynamically providing structured templates to guide users in clarifying their intentions and fostering clearer directions. This added structure serves as a scaffold for helping users frame their output evaluations with a more informed perspective on what is required in a prompt to generate a successful output, reducing the intentionality gap. While expert users can develop their own cognitive task models through LLM use to identify their own scaffolds, sharing that expertise can greatly help occasional and novice users. Furthermore, a specialized tool can guide users in specifying procedural aspects of their intentions, thereby reducing the need for repeated trial-and-error. Importantly, this prompting scaffold can change dynamically within an LLM session to better reflect the intentions as the user develops them.

As discussed earlier, a cognitive task model such as Hayes and Flowers~\cite{hayes2013new} describes how a writer moves from a goal to the executed text output. While this process is not the only viable way to generate a story, it is an approach that people often use and, therefore, may be evident in the stories captured in the textual data corpus employed in training LLMs. While the LLM was not trained to write stories, this underlying cognitive model becomes implicitly embedded in the LLM through its training set of words, phrases, and stories, and emerges in the LLM's generation performance.   While the LLM does not have an explicit cognitive model of how to generate a story, it is built through statistical analysis of a story corpus created by human writers. That is why it is \textbf{helpful for a user to mentally envision what is required for them to perform the generative task: Doing so likely engages the same task knowledge used by other writers with similar intentions.} Then, describing their developed intentions to the LLM using those cognitively informed specifications makes accessing stories with similar intent more likely through the lexical indexes built into transformer networks. 

To traverse the extreme database of connections encoded within large learning models and generate a desired outcome, a user must attempt to describe not only \textit{what} others wrote about, but also, \textit{why} they wrote. Without understanding intentions well, it is impossible to access plans from past tasks leading to good outcomes, and impossible to assess outcomes from LLM systems as meeting one's goals. \textbf{The value of generative outputs depends entirely on the intentions linking goals to their execution in output.} The three pathways shown in Figure~\ref{fig:llmmodel}  build on human knowledge of successful task completion and vary in the effort and process required. The three gaps, including the capability gap, instruction gap, and intentionality gap, together comprise the gulf of envisioning.

\section{Examining the Envisioning Gap in Three LLM Interfaces}\label{sec:cases}

Up until now, we have characterized the envisioning gulf based on the core capabilities of LLMs (i.e., language understanding and text generation) with a focus on writing tasks. We now analyze the design of three existing LLM interfaces (see Figure ~\ref{fig:case-studies}) across different generative tasks to pinpoint how the three gaps manifest during interactions. Concretely, we look at how LLM interface designs support the specific cognitive task processes they're built for -- using the framework of \textit{planning, execution, and evaluation} --  and the specific features they implement to minimize the three gaps in our revised interaction model.

\subsection{Writing using ChatGPT}
ChatGPT (Figure ~\ref{fig:case-studies}-A) is an LLM with a corresponding web-based interface developed by OpenAI~\cite{chatgpt}. Through the interactive chat interface, users can supply prompts and engage in dynamic conversations with the model. In the context of a writing task, ChatGPT can quickly produce drafts or outlines based on a given topic, elaborate on an outline, perform grammar and style checks, and paraphrase and synthesize. Naturally, the full spectrum of writing support it offers is diverse and dynamically evolving. 

\subsubsection{Capability Gap} 
To support users in understanding the action space, the ChatGPT landing page currently provides \textit{example prompts} and \textit{tasks} the model can perform. These affordances begin to reduce the capability gap during interactions. Yet the tasks depicted often lack the granularity required for users to devise concrete plans. Further, the tool also allows users to start different ``chats'' for different lines of planning and execution cycles. However, the key cognitive issue in planning is determining how to break down their goals into specific, actionable steps with the LLM. Simply having different chats may help with the organization, but it does not close the gap in helping users know how to best to formulate their intentions.

\subsubsection{Instruction gap} 
Users mainly discover how LLMs interpret prompts within ChatGPT through trial and error. These trial and error actions subsume features such as regeneration, editing the original prompt, and managing different ``chats'' with ChatGPT on the left sidebar. Yet ChatGPT does not provide a history of previous outputs, making it difficult to compare the quality of regeneration from run to run. If users cannot see how slight language changes affect outputs, they might struggle to learn from their linguistic adjustments to the prompt. In other words, the lack of feedback inhibits their ability to understand how language nuances influence the model's interpretations, making the Instruction gap even more pronounced. Separately, ChatGPT allows users to set ``custom instructions,'' which specify details and guidelines when engaging in a dialogue with the model. However, this feature requires users to foresee these intentions upfront before interacting with the model and a more general understanding of instruction utility across tasks. 

\subsubsection{Intentionality Gap}
The intentionality gap reflects the challenge users face when evaluating the LLM-generated text because they bypassed the planning and execution processes. Custom instructions are the primary mode for aligning user values with model output, acting somewhat as ``base prompts.'' There are two parts to defining custom instructions: (1) what you want ChatGPT to know about you as well as (2) how you would like ChatGPT to respond. The sum of these two mimics domain-specific prompts seen in other LLM-enabled systems, as the former asks the model to play a role in narrowing its scope while the latter helps steer the model's generation. Overall, this feature ensures that users no longer need to specify such context when prompting, allowing them to focus on crafting a good prompt based on their specific goals and intentions. However, even with this alignment, users may still encounter challenges in assessing the output because they may lack a comprehensive mental model of the content. Finally, at evaluation time, users can ask the LLM questions about its previous inputs, regenerate a response from the LLM, or even directly edit an old prompt to see how a model changes its answer. These features aim to give users more clarity on the LLM's outputs, offering ways to refine and adjust the content to better match their intentions.

\begin{figure*}[t!]
  \centering
  \includegraphics[width= 0.9\textwidth]{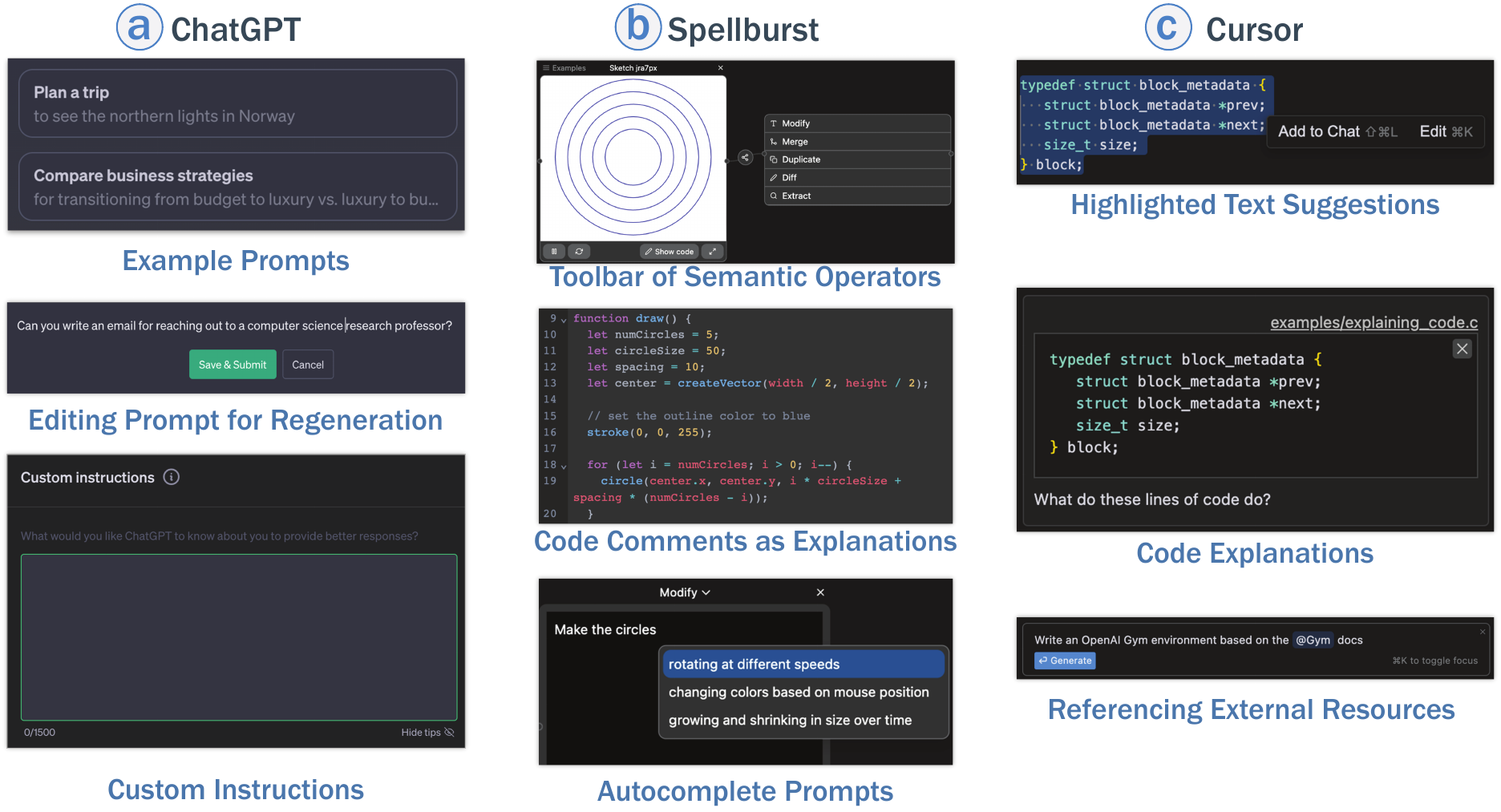}
  \caption{Example affordances for how (A.) ChatGPT~\cite{chatgpt}, (B.) Spellburst~\cite{angert2023spellburst}, and (C.) Cursor~\cite{cursor} bridges the capability (top), language (middle), and intentionality (bottom)  gaps.}
  \label{fig:case-studies}
\end{figure*}

In essence, while these features of ChatGPT offer initial guidance and organization, they don't fully equip users to carry out the planning and evaluation tasks. 

\subsection{Creative Coding using Spellburst}
Next, we look at Spellburst (Figure ~\ref{fig:case-studies}-B), a creativity support tool developed as a research artifact~\cite{angert2023spellburst}. We selected this tool as it aims to support the artists' ``exploratory creative coding'' workflow to address cognitive challenges in creative work. Specifically, they focus on bridging the artist's creative intents expressed in natural language with the implementation of those intents as code using LLM. 

Briefly, they provide a node-based interface that allows artists to prompt an LLM -- namely, ChatGPT -- to generate new computational artwork (called a sketch) and also execute creative operations such as merging different outputs or creating variations through branching. The tool provides example sketches in code to iterate from, a toolbar for each node with a set of possible operators, and autocomplete suggestions when authoring prompts. Users primarily execute the task by prompting the LLM to generate code; however, users can also manually write and edit LLM-generated code for use in further ideation. Finally, Spellburst offers various ways of evaluation, including interleaving comments in the LLM's output and allowing users to ask questions about the output through the diffing and extraction operators. The node-based canvas of the interface is also well-suited for exploring different lines of iteration after evaluation.

\subsubsection{Capability Gap} To minimize the capability gap, the interface first provides a series of example sketches. These examples, which are snippets of p5.js~\cite{p5js} code for sketches such as ``Bouncing Balls'' and ``Fractal Tree,'' provide the user with ideas for what types of tasks are best suited for the interface. In addition, the interface shortens the distance by helping users author prompts through \textit{autocomplete} based on their crowd-sourced taxonomy. This helps them understand what is possible with Spellburst, especially for users who may have an initial idea for a semantic jump. Finally, the set of semantic operators on the right side of each node, such as ``Modify,'' ``Diff,'' and ``Extract,'' showcase the different types of ways users can extend their current line of thinking using the LLM.

\subsubsection{Instruction gap} Spellburst also uses a variety of techniques to bridge the Instruction gap. After the interface is prompted to generate a sketch, the model interleaves \textit{comments} within the code of the sketch that point to the specific change in the code based on the user's prompt. These comments enable users to see how the model's inputs prompt its outputs. As an example, if a user modified a sketch to turn its outline from black to blue, the model would output the sketch with a code comment in the location where the outline color was changed. There are also numerous ways to experiment and see how the model may interpret different prompts, including the duplication and modification of a sketch node or the creation of a new branch within the interface entirely, representing a new line of semantic and syntactic jumps. Finally, two semantic operators help make the model's output more explainable by asking the LLM to describe its own output -- diffing and extracting. \textit{Diffing} allows users to see both syntactic (code-level) and semantic (prompt-based) differences between two sketches, while \textit{extracting} enables users to ask the model questions about a sketch. 

\subsubsection{Intentionality Gap} Based on their provided prompts, Spellburst asks the LLM to ``act as an expert creative coder.'' Through this approach, they are setting a high-level intentionality of how the LLM should be thinking about generating the output. Furthermore, the prompts help ensure that the model generates code that runs and compiles correctly. The team behind the system also crowd-sourced a set of image transformations to develop a creative coding taxonomy. This taxonomy drives the autocomplete suggestions, as users can orient their intentions alongside this shared vocabulary when prompting the model, improving the quality of the LLM's output.

\subsection{Software Development using Cursor}

Cursor (Figure ~\ref{fig:case-studies}-C) is a code editor that helps users pair program with AI \cite{cursor}. The system uses an LLM -- users can choose between GPT-4 \cite{openai2023gpt4} or ChatGPT \cite{chatgpt} -- to help developers chat with their project, make code changes, and address bugs. Given the growing interest in utilizing AI to supercharge the software development process \cite{copilot, 10.1145/3510454.3522684}, we analyze how Cursor implemented a human-LLM interface for the complex cognitive task process of programming.

As a fork of VSCode \cite{vscode}, Cursor has numerous tool affordances critical to supporting programming, including a file explorer, line completion, and a search bar. However, there are many features specific to Cursor. When planning, users are given several examples of tasks they can accomplish. In addition, there are various ways to converse with the LLM, including when selecting the text, using keyboard shortcuts within the main editor, and the right sidebar dedicated to chatting with the model. During execution, the user can either write code normally in the editor or prompt the model to generate code. Finally, the system provides several features for evaluation, including regenerating model responses, chatting with a model about code, and auto-debugging.

\subsubsection{Capability Gap} Cursor's first way of showcasing what tasks are available within the system is through a set of example files when a user first opens the editor. Each of these files is named after a possible task, and the comments in the file give instructions regarding what keyboard shortcuts and prompt to use to complete the goal. Similar to ChatGPT, though, since these tasks are high-level (i.e., ``fix a bug,'' ``explain code''), they are not helpful for more specific planning of goals and intentions. When interacting with code across the various panels of the interface -- including the code editor and the chat interface on the right-hand side -- the tooltip shows both what actions can be taken with an LLM and how to establish context. For instance, highlighting text within a code editor gives users the option to either add the code to the right-hand chat interface or edit directly with a prompt. Likewise, inside of the chat interface, buttons appear for users to reference a file or change the scope of the user's question within a prompt.

\subsubsection{Instruction gap} To help users understand how LLMs interpret language, Cursor allows users to ask models questions about their codebase. Topics for these questions can range from the inner workings of a particular function to the overall flow of an entire project. Since this feature can give answers about model-generated and user-written code, developers can get a better idea of how LLMs interpret their programs``under the hood.'' A similar feature that can help decipher how models understand code is prompting the model to generate inline comments. Finally, despite no direct functionality for regenerating output outside of feeding the model the same prompt, there is an option to ``rerun without context'' to see the effects of adding code or references to other materials alongside a prompt. The lack of regeneration history, however, makes it more complicated for users to map changes in their inputs to changes in model outputs.

\subsubsection{Intentionality Gap} The primary way Cursor aligns user intent with model output is through references of code snippets, files, and documentation. When authoring a prompt, users can attach an existing file of code they've written, a function within the code editor, or even documentation of third-party libraries to steer the output of the model. This functionality has many use cases, including prompting the model to adopt the same style or asking the LLM to use a particular method or function from a library. While this feature reduces the amount of effort involved in establishing context with the LLM, this affordance does not inherently give users a mental model of how the LLM works, which means that users may still run into issues. Furthermore, to help evaluate output, user can ask Cursor questions about their codebase. The questions can be about any piece of code within the code editor -- either model- or user-generated -- to help users understand how to make their prompts more optimal.

\section{Recommendations for Designing Interfaces for Envisioning}

In addition to the three interfaces described in detail in the previous section, we conducted a qualitative analysis of 12 systems to identify design patterns that would potentially support the process of envisioning. We only selected tools that were either directly accessible or at least had a video demonstration and accompanying technical description of the features, such as in research papers. For each tool, we identified affordances along the three main operators in generative tasks, namely planning, execution, and evaluation. We then clustered the identified features based on their functional similarity to develop a set of interface design patterns and corresponding tenets for teams building human-LLM systems. We do not claim a comprehensive categorization but provide a starting point for the design of interface affordances for envisioning LLM interactions.

\vspace{3mm}
\noindent
\textbf{Design Pattern 1 -- Visually Track Prompts and Outputs:}  LLM interactions are often iterative and take a trial-and-error approach. We observed that some tools provided users with a visual interface for capturing their prompting and divergent thinking through alternative pathways. A popular choice is node-based interfaces, which allow users to visualize multiple different outputs (nodes) and trace how they are connected (edges). As mentioned before, in Spellburst \cite{angert2023spellburst}, each node represents a sketch, and edges showcase different iterations of the sketch through merging, diffing, and other semantic operators. In PromptChainer~\cite{wu2022promptchainer}, a system that helps users chain together LLM prompts for complex tasks, each node represents an individual prompt and corresponding output while each edge represents the use of the output as context for another prompt. Such external representations of their thought process not only lower cognitive load but also help users better engage in more deliberate prompt authoring. For instance, users can readily see if their previous line of thinking was fruitful or make adjustments to subsequent prompts based on prior outputs. These affordances correspond to the tenet that: \textit{Users need guidance navigating the cognitive task space for prompt authoring, discerning which paths lead to desired or poor outcomes [T1].}

\vspace{3mm}
\noindent
\textbf{Design Pattern 2 -- Suggest Ideas for Prompting:} Many systems proactively offer users prompt suggestions. Some of them are aimed at assisting users who may not be as familiar with LLM-enabled interfaces, while others serve as ideation partners in the cognitive task workflow. In addition to providing ideas for prompts, examples and suggestions emphasize the importance of clarity and precision in language. If these suggestions come from examples where the model is fine-tuned, they can also help better align a user's intentions with model behavior. For example, ChatGPT~\cite{chatgpt} provides standalone prompt examples to showcase its utility. On the other hand, Notion~\cite{notion}, which is a tool for knowledge management, not only offers possible ways to use AI to improve your writing (i.e., ``Change tone''): it also gives suggestions on how to prompt an LLM to steer these improvements in a certain direction (i.e., ``Professional,'' ``Casual,'' ``Straightforward''). These patterns lead to two Tenets, namely: \textit{LLMs can serve as cognitive partners in task formulation [T2] }, and \textit{A focus on clear, precise written language will help to bridge the gap between human intention and LLM output [T3].}

\vspace{3mm}
\noindent
\textbf{Design Pattern 3 - Provide Multiple Outputs:} Rather than generate one output based on a user prompt, LLM systems may provide numerous outputs. This can be achieved by setting the \textit{temperature} of the model -- a parameter that dictates randomness -- greater than $0$ and giving the model the same prompt or explicitly asking the model to give more than one example. This feature allows users to view multiple options to see which best fits their intentions. Furthermore, providing multiple outputs helps users link the effects of changes in prompts to changes in the final output of the model. Some systems also support grouping and clustering model outputs to make this process easier. For instance, BotDesigner \cite{zamfirescu2023johnny} lets users manually assign a tag to model outputs, while Sensecape \cite{suh2023sensecape} groups relevant topics together semantically based on a high-level topic. These features support the tenet that \textit{ LLMs should support users through their divergent thinking strategies [T4].}

\vspace{3mm}
\noindent
\textbf{Design Pattern 4 - Make the Output Explainable:} Some systems prompt LLMs to explain their outputs or make them more interpretable. This design pattern allows users to better understand how LLMs interpret certain prompts and makes model outputs easier to use for manual editing. How this technique is applied in practice can differ depending on the task domain. Replit~\cite{replit}, a browser-based code editor, has an AI assistant named Ghostwriter that generates in-line comments within its code responses. Another code editor, Cursor~\cite{cursor}, does not always provide code comments but does allow users to ask LLMs about the code they generate. In contrast, Sensecape \cite{suh2023sensecape}, which is designed for exploration and sensemaking, prompts an LLM to return a response at different levels of detail, such as through summaries and keywords. These features help users address their intentionality gaps and better assess the model output. This pattern supports the tenet \textit{An error in human-LLM interaction is not just a user error or LLM failure but signals a breakdown in the distributed cognitive system that requires collaborative repair [T5].}  \change{However, in designing for explainability and drawing causal inferences between prompt inputs and outputs, design should account for users' overreliance on explanations without careful validation~\cite{fok2023search}}.

\vspace{3mm}
\noindent
\textbf{Design Pattern 5 - Use domain-specific prompting strategies:} Outside of standard prompt engineering techniques, most systems use a custom prompting strategy depending on their task. These methods help steer the outputs from LLMs into something usable for the end goal while also minimizing the output ambiguity that may arise in trying different prompts. As an example, Graphologue~\cite{jiang2023graphologue}, which is designed to turn text-based responses from LLMs into graphical diagrams, uses prompting techniques to have models annotate entities and relationships within their outputs to create diagrams in real-time. Coding Steps~\cite{kazemitabaar2023studying}, a web-based application to help novices learn Python, prompts models with static examples, then user code, then the user prompt, to ensure that the level of output is appropriate for beginners. These strategies allow designers to implement conceptual tasks for users and consequently allow them to build task-specific system mental models. The corresponding tenet is that, \textit{Users favor working with a system mental model leading to actions when working within a defined task domain [T6].}

\vspace{3mm}
\noindent
\textbf{Design Pattern 6 - Allow manual control of output}: Many systems afford users the opportunity to manually edit the outputs and interactions with LLMs. Since many LLM-enabled systems are built for exploration and ideation, direct manipulation can help users better incorporate their values and intentions into the model. Oftentimes, manual editing is introduced when one output serves as input to another LLM. For instance, while LIDA~\cite{dibia2023lida}, a tool for generating visualizations and infographics, prompts an LLM to output goals for dataset exploration, users are also allowed to enter their own goals and adjust the model's suggestions. Likewise, Mirror~\cite{xu2023mirror} -- an NLI for data querying and summarization -- gives users the ability to edit the SQL queries generated by a pre-trained code model to add human expertise. These features align with the tenet, \textit{If tasks are well-defined, people prefer dedicated interfaces over dynamic interfaces [T7].}
\section{Discussion}\label{sec:discussion} 

\change{In this work, we have theorized about cognitive challenges emerging in the transition from conventional software paradigms to prompt-based interactions powered by generative models. Based on prior empirical evidence on challenges with prompting~\cite{zamfirescu2023herding,zamfirescu2023johnny,kim2023understanding}, we have applied cognitive science and HCI perspectives to characterize significant HCI design challenges with prompt-based interactions. Given the advanced cognitive capabilities of LLMs, people are now able to express in natural language their bespoke task goals and ask the LLM to perform those goals for them. At the same time, they lack the specific affordances of conventional systems in formulating their intentions and task plans and evaluating the LLM outputs. Given the shift in the operational scope from deterministic functions to dynamic intelligent agents, we have identified new cognitive process models for specifying actions through intentions, i.e., the process of envisioning. In reasoning about envisioning intentions with LLMs, we have also identified three specific gaps including the capability gap, instruction gap, and intentionality gap, and we have provided initial recommendations for interface designers to scaffold prompting. However, a number of open questions remain about designing prompt-based interfaces. Here, we propose open questions for future research as we consider future development and applications of generative models.}

\subsection{Open Questions for Designing Human-LLM Interactions}

\subsubsection{How should we model conversational interactions between humans and LLMs?}
\change{Given the human-like conversational abilities of LLM interactions, designers must consider how to effectively model conversations. Recent work on designing natural language interfaces (e.g.,~\cite{setlur2022you,dombi2022common} ) has primarily considered the `Recipient Design' approach based on Grice's Cooperative principles~\cite{grice1975logic,stone2005communicative,branigan2010linguistic}. This approach models communication as sensitive to context and individual characteristics, and it recognizes that speakers tailor their speech and communicative behavior to meet the needs of their listeners. As the influential literary theorist Mikhail Bahktin theorizes ~\cite{bakhtin2010dialogic}, any utterance has both addressability (every word always addressed to someone)  and answerability (every word directed toward and anticipating an answer from someone). Sociolinguists describe these features as manifest in the \textit{recipient design} of an utterance ~\cite{goodwin1990conversation}.   The maxims of quantity, quality, relation, and manner guide how information is conveyed and interpreted to ensure clarity, relevance, and truthfulness. However, with LLM dialog, such an inductive approach can be challenging for users as these maxims depend on each conversationalist tracking a theory of mind~\cite{frith2005theory} regarding what their partner thinks and knows. An AI capable of a complex theory of mind for individual users across sessions will likely require extensive development and testing for feasibility. An alternative model is the Transactional Model of Communication~\cite{sameroff2009transactional}, which offers a more dynamic view of interchange. Rather than precision, a deductive process allows for repairs and adjustments of misunderstanding and miscommunication, and consider the contextual and continuous nature of communication. Applying this model for human-LLM interaction, rather than emphasizing the quality of communication, in interfaces offers opportunities for repair and feedback, e.g., editing a previous prompt in the ChatGPT interface.}

\change{Other models of communication, such as the Socio-Cognitive approach to Pragmatics~\cite{kecskes2010paradox} and Speech Act Theory~\cite{searle1980speech}, may also be useful for design. The socio-cognitive approach accounts for how the speaker's and listener's background, intentions, and situational contexts contribute to constructing meaning. This approach may provide guidance for characterizing the agentic roles of LLMs in personalization and adaptability, modeling intentions and expectations, code-switching, and ethical and responsible interactions. Further, focusing on the `functions' of language, i.e., speech acts, may allow design for conversational rules and conventions. Lastly, according to social exchange theory~\cite{cropanzano2005social}, communication behaviors are influenced by design to maximize benefits and minimize the cost of interactions.  Future research should investigate the balance and trade-offs between high-quality intuitive prompting and the adaptability offered by more deductive conversational approaches. This exploration should focus on how varying models of communication can enhance LLMs' ability to understand and adapt to diverse user backgrounds, intentions, and contexts. By integrating insights from these communication models, HCI research can develop LLM interfaces that not only facilitate efficient exchanges but also support ethical, personalized, and contextually sensitive interactions.}

\subsubsection{What is a useful theory of mind for LLMs to support effective envisioning?} 
\change{In conventional systems, users' mental models of systems provide a functional account of how the system produces its output, and that knowledge is used to generate necessary action specifications and evaluate whether the output is `` good enough.'' In the case of LLMs, novice users likely do not have a system mental model that can sufficiently describe what happens at a process level, or that understanding does not allow the prediction of output from input. The requisite mental model to account for an LLM's performance includes both the training process and the algorithms used to learn; further, it must include a theory of \textit{what} was learned by the LLM. If we consider LLMs from a purely human-like perspective, as in human-human communication, we require a cognitive understanding of how communication works, including forming assumptions and expectations,  norms, shared knowledge, feedback processing, symbolic understanding, perspective taking, reciprocity and feedback processing, etc. For another human, I can use my own mind to generate predictions of what output will likely be produced by another mind using my own mind as a guide; for example, the colors red, white, and blue make me think of America. However, no other minds have the equivalent dataset of an LLM. Its scope defies the predictions one can generate from the information encoded in one human mind. Without a sense of the outcome of its learning process, it is impossible for another mind to predict LLM output. That is why LLMs are exciting generators different from humans but connected through text descriptions of human experiences. It is sufficient to engage in a conversational LLM interaction and feel like talking with another person who has a theory of mind about the individual and associated beliefs, feelings, and goals through the power of intentions.}

\change{Future research should explore how to bridge the gap between the complex, often opaque inner workings of LLMs and the intuitive understanding of its users. This work will involve creating models that accurately represent the LLM's operations in a user-friendly manner, aligning them with the system's actual functionality. The challenge lies in simplifying the complex mechanisms of LLMs without sacrificing essential details that users need to predict the outcomes of their interactions. Prior research on end-user programming has studied the challenges novices face in envisioning simple interactive features and ways to support requirement specification, debugging, and verification~\cite{ko2004six}. Similarly, for LLMs, theory should account for the evolving nature of user requirements and the emergent design process, acknowledging that users often learn and refine their understanding of LLMs through experience. Efforts should be concentrated on enhancing the visibility of the LLM's processing pathways, perhaps through interactive visualizations or simplified explanatory frameworks to reinforce user understanding through better prompting, iteration, and evaluation.}

\subsubsection{What is the optimal ``sweet spot'' for Human-LLM interaction along the three interaction dimensions?}

\change{Reconsidering the three dimensions in Figure~\ref{fig:teaser}, LLMs are appealing because the inputs can be abstract, complex, and vague (i.e., underspecified), and they can still produce outputs that are good. As mentioned earlier, if the quality of the answer matters, end users will want to iteratively explore the generative features to reach a high-quality answer. In such cases, envisioning can be challenging, and in both our revised model and as seen in prior work~\cite{zamfirescu2023herding}, we need better guidance in designing interfaces to make envisioning easier for users. When powering interfaces with LLM capabilities, how do we balance Intent Specificity, Functional Flexibility, and Output Determinacy in ways that leverage the strengths of the LLM while aligning with the user's needs for quality and relevance? For Intent Specificity, a semi-structured (templated) approach is advantageous. While LLMs excel at interpreting and generating responses to open-ended queries, a certain level of specificity in the user's intent can guide the LLM to produce more targeted and pertinent content. Functional Flexibility should be oriented towards dynamic responsiveness. The LLM's ability to pivot across different tasks and domains should be fully utilized within a framework that is shaped by the context of the interaction. Future research should investigate efficient pathways and optimal points along these three dimensions. Such inquiries will be likely to include novel interfaces that suggest modifications to vague inputs and feedback systems that learn from each interaction to refine future responses. Moreover, future research should also consider the evolving nature of user expectations and the continuous advancements in LLM capabilities, ensuring that the ``sweet spot'' for AI and interface design remains a moving target that adapts to the growing sophistication of both users and technology.}

\subsection{Limitations and Future Work}
Of course, LLMs have been game-changing for AI, and have launched a wildly diverging portfolio of applications. They evidence the fact that much of everyday human task performance is rote, standardized, and repetitive. Strong patterns across the text database reveal how little originality exists in the accumulated human text products available in digital form. It is quite possible that the deep cognitive processing proposed here as a means of probing LLMs to produce output more similar to desired human outcomes is rare. However, its value when it does occur suggests it is well worth developing models of the cognitive processes shared across cognitive tasks.

Another limitation of our model is the obvious differences among people in their ability to identify intentions from goals. Divergent thinking ~\cite{guilford1956structure} is a minor part of academic training across the school curriculum, whereas converging on a single correct answer dominates learning. People show major differences in their ability to solve open-ended problems and complete generative tasks, potentially attributable to differences in cognitive capacities, including memory but also imagination. A common test of creativity, the unusual uses test, asks for different ways to use a common object, such as a brick~\cite{guilford1956structure}. People often fixate on functions such as using the brick as a weight, and generate uses like an anchor, paperweight, and balloon holder. Less often, they identify unusual functions, such as using its material as a dye for crayons or lipstick. Perhaps using human minds as the ``key'' limits solutions to only those generated by the human mind. While the LLMs are currently making use of the text-based products of human minds, it is possible that future systems will encode different forms of data less dependent on human goals and intentions and more content generated by AI models. Further, more systems with corpora combining products of humans and AIs may diverge further away from cognitive models as explanations of links between input and output.

Finally, this work is intended to propose a direction for the development of new approaches to HCI. Further pathways for human-LLM interaction can be identified, and new interface supports for LLM use based on prompting guidance are growing daily. While other approaches may be quicker (and dirtier) so as to plug obvious holes in current LLMs, the promise of this work is to capture the intentions found useful by humans in executing generative tasks.  To test this approach, comparing prompts where the intention is evident will determine its value in creating satisfying outputs from LLMs. Strategies from co-work, such as asking someone to repeat back their understanding of the task instructions given, may prove similarly useful with LLMs. An empirical agenda can determine not just the factual or writing quality of AI systems but also their value to human users. This proposed approach to HCI with LLMs aims to support the user as they must think more deeply and fully during interactions with systems in order to integrate their processing abilities with the strengths offered by systems. This work, like the development of LLMs, is at its beginning stages.

\section{Related Work}

\subsection{Documented Challenges of LLMs} \label{sec:llmchallenges}
A core challenge of using LLMs is their \textit{explainability}~\cite{mathews2019explainable}, i.e., how can we explain why a model behaves in a certain manner? Compared to the domain of traditional machine learning, LLM explainability is a different challenge~\cite{zhao2023explainability}, as these models are pre-trained to do a variety of complex reasoning tasks~\cite{yang2023harnessing} and absorb patterns from data automatically~\cite{mirchandani2023large}. Regarding inputs to LLMs (i.e., prompts), the largest issue is that it is not always clear how a prompt strategy affects model output~\cite{liu2021pretrain, sanh2022multitask}. Even for popular methods like chain-of-thought -- which asks a model to explain itself -- there is no evidence to suggest whether models reason towards the answer through the steps they provide, based on their pre-training data, or through other heuristics~\cite{saparov2022language}. Overall, explainability, or the lack thereof, is a significant contributor to the gaps involved in LLM interaction, as users struggle to build a mental model~\cite{bhatt2021uncertainty, sun2022investigating, vasconcelos2023generation}. 

Another challenge with LLMs is concerned with the \textit{usability} of their outputs. For instance, such models can hallucinate, where the text generated seems structurally correct but is actually nonsensical or incoherent~\cite{10.1145/3571730, 10.1145/3442188.3445922, rawte2023survey}. In addition, LLMs do not always produce factually correct output, and it may be difficult to verify whether the output is correct or not~\cite{ji2023survey, maynez-etal-2020-faithfulness}. Depending on the context and domain, such as in the realms of medical or military applications~\cite{oniani2023military}, this inaccuracy can be severely detrimental~\cite{koga2023exploring, sallam2023chatgpt, lee2023benefits}. Furthermore, while a prevalent technique for addressing this shortcoming is to provide sources, it can be hard to implement in practice and may also not always be correct~\cite{52046, liu2023evaluating, rashkin2023measuring}. Combined with the issue of explainability, it can be difficult to correct and steer LLMs to responses that are more usable.

Lastly, a final challenge concerns the issue of \textit{bias} in LLMs. Much empirical research has shown that a plethora of biases are encoded in these models, including racial bias, gender bias, and bias around political leanings, to name but a few~\cite{sheng-etal-2019-woman, nadeem2020stereoset, talboy2023challenging, feng2023pretraining}. There is also an abundance of work in the realm of jailbreaking LLMs to generate toxic outputs and leak private information in both their training data and conversation history~\cite{wang2023decodingtrust, casper2023explore, liu2023jailbreaking}. There are many factors contributing to this propagation of bias~\cite{ferrara2023should}, including the corpora these models are trained on, how the data is labeled and annotated, and the architectures that power these LLMs under the hood~\cite{doi:10.1126/science.aal4230, 10.1145/3442188.3445922, munro-etal-2010-crowdsourcing}. Overall, while existing models reject most harmful prompts (i.e., ChatGPT responds with ``I'm sorry, but I can't assist with that request.''), socially situated contexts can still produce potentially offensive output from an LLM~\cite{shaikh-etal-2023-second}.

\subsection{Prompt Engineering} \label{sec:promptengg}
Prompt engineering encompasses the set of techniques used to converse with LLMs. These methods assist with setting rules, structuring output, and overall guiding the model in the direction in which a user intends~\cite{white2023prompt}. While there has been much research devoted to uncovering these emergent properties, these techniques are often simple tricks that are intended to mimic the process of human reasoning.

One of the most effective prompting strategies is providing examples of expected input and output, also known as few-shot prompting~\cite{brown2020language}. The inspiration comes from human cognition, as people can learn new concepts from a small set of examples while also applying these concepts to new inputs~\cite{lake2019human, lake2016building}. The effects of few-shot prompting are more pronounced when models are of a certain scale, and there are numerous factors that can aid or inhibit the helpfulness of such prompts~\cite{kaplan2020scaling}. These include the semantic similarity of the training examples to the test examples, the choice of prompt format, and even the order of the examples in the prompt~\cite{liu2021makes, zhao2021calibrate}.

Another predominant technique used in prompting is chain-of-thought, or providing a series of reasoning steps to show the model how to get to the final answer~\cite{NEURIPS2022_9d560961, kojima2023large}. These prompts are helpful to learning because, for humans, explanations break down why a certain answer is correct and not just what the final answer is~\cite{lampinen2022language, Ahn_1992, LOMBROZO2006167, prystawski2023psychologicallyinformed}. Coupled with few-shot prompting, there are numerous factors that make for useful chain-of-thought prompts, such as the amount of complexity within the prompts (measured by the number of steps), whether the provided examples are relevant to future queries, and the correct ordering of the reasoning steps~\cite{fu2023complexitybased, wang2023understanding, chen2023need}. There are also a plethora of variations that build upon chain-of-thought, including sampling multiple responses given the same prompt~\cite{wang2023selfconsistency, Stanovich_2000} (Self-Constency Sampling); repeatedly prompting a language model to ask follow-up questions~\cite{press2023measuring} (Self-Ask); and both decomposing a problem into numerous steps and sampling numerous responses at each step \cite{yao2023tree} (Tree of Thoughts).

\subsection{Designing Human-AI Interfaces}
\label{sec:designnli}
The rise of artificial intelligence (AI) has led to a surge of interest in the development of human-AI systems. The creation of these novel systems has brought new challenges. For instance, since these models are sometimes perceived as non-deterministic ``black-boxes,'' users can have a hard time discerning how these interfaces produce their outputs~\cite{hoffman2018explaining, bathaee2017artificial}. Likewise, given the massive amount of data these models are trained on, there are new concerns around the bias and privacy of such systems~\cite{bolukbasi2016man, kostick2022mitigating, yeom2018privacy}. Perhaps most critical is the change in the relationship between humans and interfaces: while traditional NLI systems served more as assistants to end users~\cite{farooq2016human, wickens2015engineering}, human-AI systems act more as collaborators due to their human-like cognitive abilities~\cite{lyons2018viewing, brandt2018human}. As a result of this role change, there are many propositions for human-AI interaction guidelines~\cite{amershi2019guidelines, mohseni2021multidisciplinary, wright2020comparative}, including conveying the consequences of user actions, providing diverse options from models, and holding the system accountable for errors. \change{Further prior research has looked at changing design practices to accommodate the new challenges of designing user experiences for machine learning capabilities~\cite{yang2018investigating,yang2020re,subramonyam2021towards,subramonyam2022solving}.}

\change{Specific to natural language interfaces, prior work has looked at designing natural language interfaces and LLM-powered chatbots~\cite{yang2019sketching,zamfirescu2023herding}.} These interfaces are most prevalent in the domains of data visualization~\cite{9118800, 10.1145/2807442.2807478} and querying~\cite{affolter2019comparative, Hozcan2020state}. There are several issues involved in the development of NLI systems, including the ambiguity of natural language, communicating to users what the system can do, and evaluating the utility of these interfaces in accomplishing their end goal~\cite{shen2022towards}. To this end, researchers within academia and industry have put forth ideas for creating effective NLI systems, including the use of autocomplete to show users how to phrase their queries to the system~\cite{bacci2020inspecting, yu2019flowsense}, the design of conversational interfaces to engage users in a back-and-forth dialogue about their intentions~\cite{setlur2022you, setlur2016eviza, 8019833}, and the development of multi-modal features to give users more control over their input to the interface~\cite{kassel2018valletto, srinivasan2020interweaving}.

Finally, recent work in HCI has been focused on generative AI, in particular, LLMs~\cite{chatgpt, bard, claude} and diffusion models~\cite{ramesh2022hierarchical, rombach2022high}. Within this space, significant effort has been concentrated on developing effective prompt strategies (see Section ~\ref{sec:promptengg}), helping users craft and author these prompts~\cite{zamfirescu2023johnny, 10.1145/3544549.3585737, wu2022ai}, and also developing preliminary guidelines for generative models~\cite{chen2023next, weisz2023toward}. Focusing specifically on LLMs, though, what is missing in the current literature is a better understanding of \textit{why human-LLM interaction is different from human-AI interaction} -- and more broadly, \textit{how human-LLM systems are different from traditional natural language interfaces} -- and what types of strategies designers can employ to address the gaps that result from these new computational opportunities.

\section{Conclusion}
Our work applies a cognitive framework to characterize the dynamics of prompt-based interfaces such as ChatGPT, highlighting the complexities of interacting with LLMs. While LLMs are capable of interpreting a vast range of queries, their very flexibility can pose challenges for users attempting to convey precise intentions. We identify and characterize a new kind of interaction gulf called the ``gulf of envisioning,'' which captures the challenge users face in successfully formulating their intentions to elicit the desired response from an LLM.  This gulf is further identified by the capability gap -- what intentions can the LLM perform, the instruction gap --how to say what is needed to the LLM, and the intentionality gap -- what to expect and how to evaluate the generated output, all of which describe varying facets of human-LLM misalignments. By arguing that for LLM interfaces, ``intentions are actions,'' we provide design recommendations to support the process of envisioning with generative AI models.

\begin{acks}
We thank the reviewers for their feedback on the paper. Subramonyam and Agrawala are supported through the AI Research Institutes program by the National Science Foundation and the Institute of Education Sciences, U.S. Department of Education through Award $\#2229873$ - National AI Institute for Exceptional Education. Any opinions, findings and conclusions or recommendations expressed in this material are those of the author(s) and do not necessarily reflect the views of the National Science Foundation, the Institute of Education Sciences, or the U.S. Department of Education.
\end{acks}





\bibliographystyle{ACM-Reference-Format}
\bibliography{99_refs}

\end{document}